\documentclass[10pt]{article}
\usepackage[cp1251]{inputenc}
\usepackage[dvips]{epsfig}   \usepackage[dvips]{changebar}
\usepackage[dvips]{graphicx}
\usepackage{cite}

\def\lsim{\
  \lower-1.2pt\vbox{\hbox{\rlap{$<$}\lower5pt\vbox{\hbox{$\sim$}}}}\ }
\def\gsim{\
  \lower-1.2pt\vbox{\hbox{\rlap{$>$}\lower5pt\vbox{\hbox{$\sim$}}}}\ }

\usepackage{amsmath,enumerate, amsfonts, amssymb,amsthm}

\begin{document}
\title{Dispersion law for a one-dimensional weakly interacting Bose
gas\\ with zero boundary conditions}
 \author{Maksim Tomchenko
\bigskip \\ {\small Bogolyubov Institute for Theoretical Physics} \\
 {\small 14b, Metrolohichna Str., Kyiv 03143, Ukraine} \\
 {\small E-mail:mtomchenko@bitp.kyiv.ua}}
 \date{\empty}
 \maketitle
 \large
 \sloppy
\textit{From the time-dependent Gross equation, we find the
quasiparticle dispersion law  for a one-dimensional weakly
interacting Bose gas with a non-point interatomic potential and zero
boundary conditions (BCs). The result coincides with the dispersion
law for periodic BCs, i.e. the Bogolyubov law $E_{B}(k) =
\sqrt{\left (\frac{\hbar^{2} k^2}{2m}\right )^{2} +
n_{0}\nu(k)\frac{\hbar^2 k^2}{m}}$. In the case of periodic BCs, the
dispersion law can be easily derived from Gross' equation. However,
for zero BCs, the analysis is not so simple. } \\

 \section{Introduction}
 The quasiparticle dispersion law is usually derived
under periodic boundary conditions (BCs), because under such BCs the
analysis is the simplest
\cite{bog1947,bz1955,gross1963,lieb1963,yuv2,zagrebnov2001}. In
nature, however, BCs are typically close to zero ones. An analysis
under zero BCs is of independent interest from a mathematical point
of view. From a physical point of view,  it is interesting to
ascertain whether the boundaries affect the dispersion law $E(k)$ of
the quasiparticles. Another interesting point is that under zero
BCs, the wave functions of the system are not eigenfunctions of the
momentum operator. Therefore, the value of $k$ in the quasiparticle
dispersion law $E(k)$ is a quasimomentum (instead of momentum).

The dispersion law $E(k)$ for a one-dimensional (1D) system of
weakly interacting bosons with zero BCs  has already been found
using the Bethe ansatz \cite{mtsp2019} (for a point interaction) and
by a generalization of Bogoliubov's method \cite{mtmethodbog} (for a
non-point interaction). Both solutions coincide with the Bogoliubov
law \cite{bog1947,bz1955}
\begin{equation}
E_{B}(k) = \sqrt{\left (\frac{\hbar^{2} k^2}{2m}\right )^{2} +
n_{0}\nu(k)\frac{\hbar^2 k^2}{m}}.
     \label{0} \end{equation}
It was also found within Haldane's harmonic-fluid approach that the
sound velocity in a 1D Bose system with zero BCs is the same as that
in a similar system with periodic BCs \cite{cazalilla2004}. The
dispersion law (\ref{0}) is in approximate agreement with the
experiment
\cite{ketterle1997,ketterle1999,steinhauer2002,ozeri2002,ozeri2005,steinhauer2012,nagerl2015,hadz2021}
(true, all experiments except \cite{hadz2021} were performed with a
nonuniform gas in a trap). The results in \cite{mtsp2019} hint that
in the case of zero BCs, quasimomentum seems to be an additional
(non-additive) integral of motion instead of an additive momentum.
However, we do not know whether there exists an operator that
corresponds to the quasimomentum of the system and commutes with the
Hamiltonian of the system.

In the present work, we  determine the dispersion law of a 1D system
of weakly interacting bosons under zero BCs and in the absence of an
external field. The interatomic potential is assumed to be in its
general form. We  use Gross' approach
\cite{gross1963,gross1957,gross1958} which is the simplest method of
describing a weakly interacting Bose system. In this case, we obtain
new solutions of Gross' equation (\ref{1}). Such an approach
complements the methods in \cite{mtsp2019,mtmethodbog,cazalilla2004}
and helps to better understand the properties of quasiparticles
under zero BCs, including the nature of the quasimomentum of the
quasiparticle.

\section{Solutions of Gross' equation}
 For simplicity, we restrict ourselves to a 1D case.
Consider a weakly interacting Bose gas placed in a vessel with zero
BCs in the absence of an external field. Let the temperature be
extremely low, $T\rightarrow 0~\mathrm{K}$. Such a gas can be
described by the time-dependent Gross equation
\cite{gross1963,gross1958}
\begin{eqnarray}
i\hbar\frac{\partial \Psi(x,t)}{\partial t} =
-\frac{\hbar^2}{2m}\frac{\partial^{2} \Psi(x,t)}{\partial x^{2}}
 + \Psi(x,t) \int\limits_{0}^{L} d x^{\prime} U(|x-x^{\prime}|)|\Psi(x^{\prime},t)|^{2}
     \label{1} \end{eqnarray}
with the normalization
\begin{equation}
 \int\limits_{0}^{L} d x |\Psi(x,t)|^{2}=N,
     \label{2} \end{equation}
where $N$ is the total number of particles. Let the system be in
interval $x\in\lbrack 0,L]$. Zero BCs mean that %
\begin{equation}
\Psi(x=0,t)=\Psi(x=L,t)=0. \label{3}%
\end{equation}
Gross' equation (\ref{1}) follows from the Heisenberg operator
equation if we set $ \hat{\psi}(x,t) =\Psi(x,t)$ in the latter
\cite{gross1963,gross1958}. In this case, $N$ must be large ($N\gg
1$ \cite{bog1947}), and $\Psi(x,t)$ can be regarded as the wave
function (WF) of a nonuniform quasicondensate. Later, equation
(\ref{1}) was derived by three other methods
\cite{esryphd,salasnich2000,gp1}.

If we replace the interatomic potential $U(|x-x^{\prime}|)$ by a
point-like one $2c\delta(x-x^{\prime})$, then Eq. (\ref{1}) passes
into the Gross--Pitaevskii equation
\cite{pit1961,gross1961} %
\begin{equation}
  i\hbar\frac{\partial\Psi(x,t)}{\partial t}=-\frac
{\hbar^{2}}{2m}\triangle \Psi(x,t) +  2c|\Psi(x,t)|^{2}\Psi(x,t).
\label{4} \end{equation}%
In what follows we consider the general case of a non-point
potential $U(|x-x^{\prime}|)$.

Note that  condensate $N_{k=0}\sim N$ for an \textit{infinite} 1D
uniform system is forbidden at $T> 0$
\cite{bogquasi1,bogquasi2,mermin1966,kane1967,reatto1967,hoh1967,pethick2008}
and even at $T = 0$
\cite{pethick2008,lenard1964,popov1972,popov1980,schwartz1977,vaidya1979,haldane1981,petrov2004,mt2016}.
However, all systems in nature are finite. For a finite 1D uniform
system at $T\geq 0$, the lowest macroscopically occupied state is
permitted, although its properties correspond to a quasicondensate
\cite{pethick2008,petrov2004} rather than a condensate. For our
approach, the distinction between condensate and  quasicondensate is
inessential; both can be described by  WF $\Psi(x,t)$.

The quasicondensate WF reads
\begin{equation}
\Psi(x,t)=R(x,t)e^{iS(x,t)/\hbar}, \label{5}%
\end{equation}
where $R$ and $S$ are real functions. Let the quasicondensate in the
ground state be described by the WF
\begin{equation}
 \Psi_{0}(x,t)  = R_{0}(x) e^{i S/\hbar}, \qquad S=-\epsilon t,
     \label{6} \end{equation}
which satisfies the equation
 \begin{eqnarray}
\epsilon \Psi_{0} = -\frac{\hbar^2}{2m}\frac{\partial^{2}
\Psi_{0}}{\partial x^{2}}
 + \Psi_{0}(x,t) \int\limits_{0}^{L}  d x^{\prime} U(|x-x^{\prime}|)|\Psi_{0}(x^{\prime},t)|^{2}.
     \label{7} \end{eqnarray}
The BCs are $R_{0}(x=0,t)=R_{0}(x=L,t)=0$. Such a quasicondensate is
uniform everywhere except in a very narrow region near the walls,
and the particle number density is $n_{0}(x)=R_{0}^{2}(x)$. In the
presence of small oscillations, we obtain:
\begin{equation}
 R(x,t)=R_{0}(x)+\delta R(x,t), \qquad S(x,t) =-\epsilon t+s_{0}(x,t),
     \label{8} \end{equation}
and $n(x)\equiv R^{2}(x) =n_{0}(x)+\tilde{n}_{0}(x,t)$. Because $R$,
$n$ and $S$ are real, the quantities $\tilde{n}_{0}(x,t)$ and
$s_{0}(x,t)$ must also be real. By substituting function (\ref{5})
with $R(x,t), S(x,t)$ (\ref{8}) into Eq. (\ref{1}) and separating
the real and imaginary parts, we obtain the following equations for
the small quantities $\tilde{n}_{0}(x,t)$ and $s_{0}(x,t)$:
\begin{equation}
 \frac{\partial \tilde{n}_{0}}{\partial t}  = -\frac{1}{m}\nabla\left [(n_{0}+\tilde{n}_{0}) \nabla s_{0} \right ],
      \label{7b} \end{equation}
\begin{eqnarray}
 -\frac{\partial s_{0}}{\partial t} & =& -\frac{\hbar^{2}}{2m}\frac{\nabla^{2}
 [(n_{0}+\tilde{n}_{0})^{1/2}-n_{0}^{1/2}]}{(n_{0}+\tilde{n}_{0})^{1/2}}
 + \frac{(\nabla s_{0})^{2}}{2m} \nonumber \\ &+&
 \frac{\hbar^{2}}{2m}\frac{\delta R}{R_{0}}\frac{\nabla^{2}R_{0}}{R_{0}+\delta R}
 +\int\limits_{0}^{L} d x^{\prime}\tilde{n}_{0}(x^{\prime},t)U(|x-x^{\prime}|).
      \label{8b} \end{eqnarray}
To simplify these equations, we neglect the  nonuniformity of $
R_{0}(x)$ near the walls by setting $\nabla n_{0}=0$ and $\nabla
R_{0}=0$. In this case $n_{0}(x)=n_{0}=N/L$. If $N\gg 1$ and the
coupling is not too weak, the values of $\nabla n_{0}$ and $\nabla
R_{0}$ are significantly different from zero only at a distance from
the wall, which is less than or of the order of the mean interatomic
distance $\bar{R}=L/N$ \cite{gp1}. Therefore, we may expect that
consideration of the nonuniformity of $ R_{0}(x)$ near the wall will
affect only the solution for the ground state, but not the
dispersion law, because the latter is a bulk property.

Furthermore, we consider the oscillations of the density and the
phase to be so weak that their smallness exceeds the smallness of
the potential. In this case we can restrict ourselves by a linear
approximation. Thus, we obtain \cite{gross1963}
\begin{equation}
 \frac{\partial \tilde{n}_{0}}{\partial t}  =  -\frac{n_{0}}{m}\nabla^{2}s_{0},
      \label{9} \end{equation}
\begin{equation}
 -\frac{\partial s_{0}}{\partial t}  = -\frac{\hbar^{2}}{4mn_{0}}\nabla^{2} \tilde{n}_{0}
  + \int\limits_{0}^{L} d
  x^{\prime}\tilde{n}_{0}(x^{\prime},t)U(|x-x^{\prime}|),
      \label{10} \end{equation}
and the zero BCs take the form
\begin{equation}
 \tilde{n}_{0}(x=0,t)=0, \quad \tilde{n}_{0}(x=L, t)=0.
\label{10bc} \end{equation}

Because only standing waves can be stationary in the presence of
boundaries, we seek  solutions  in the form
\begin{equation}
 \tilde{n}_{0}(x,t)= \tilde{n}(x)T_{\rm{n}}(t), \qquad s_{0}(x,t)= s(x)T_{\rm{s}}(t).
      \label{11} \end{equation}
Substituting these functions into Eqs. (\ref{9}) and (\ref{10}) and
separating the variables, we obtain:
\begin{equation}
 \frac{1}{T_{\rm{s}}(t)}\frac{\partial T_{\rm{n}}(t)}{\partial t}  =  C_{1},
      \label{12} \end{equation}
\begin{equation}
 -\frac{1}{T_{\rm{n}}(t)}\frac{\partial T_{\rm{s}}(t)}{\partial t}  =  C_{2},
      \label{13} \end{equation}
\begin{equation}
 C_{1}\tilde{n}(x)  =  -\frac{n_{0}}{m}\nabla^{2}s(x),
      \label{14} \end{equation}
\begin{equation}
 C_{2} s(x)  = -\frac{\hbar^{2}}{4mn_{0}}\nabla^{2} \tilde{n}(x)
  + \int\limits_{0}^{L}  d x^{\prime}\tilde{n}(x^{\prime})U(|x-x^{\prime}|).
      \label{15} \end{equation}
The solution of Eqs. (\ref{12}) and (\ref{13}) can be written as
\begin{equation}
 T_{\rm{n}}(t)  =  \cos{\omega t}, \qquad   T_{\rm{s}}(t)  =  \sin{\omega
 t}, \qquad C_{1}  =  C_{2} = -\omega
            \label{17} \end{equation}
in the real form or
\begin{equation}
 T_{\rm{n}}(t)  =  T_{\rm{s}}(t)  =  e^{i \omega t}, \qquad
 C_{1}  = - C_{2} =  i\omega
      \label{19} \end{equation}
in the complex form. The real values of  $\tilde{n}_{0}(x,t)$ and
$s_{0}(x,t)$ are only obtained for the solution (\ref{17}), which is
used in the following. Then the equations for $\tilde{n}(x)$ and
$s(x)$ take the form:
\begin{equation}
\omega\tilde{n}(x)  =  \frac{n_{0}}{m}\nabla^{2}s(x),
      \label{20} \end{equation}
\begin{equation}
 -\omega s(x)  = -\frac{\hbar^{2}}{4mn_{0}}\nabla^{2} \tilde{n}(x)
  + \int\limits_{0}^{L} d x^{\prime}\tilde{n}(x^{\prime})U(|x-x^{\prime}|).
      \label{21} \end{equation}
These are the two basic equations that we will study here. They are
simple enough but are generally not easily solvable.

Let us try to find a solution as a single harmonic:
\begin{equation}
\tilde{n}(x)=n_{0} a_{2l} \sin{(k_{2l}x)}, \qquad
s(x)=b_{2l}\sin{(k_{2l}x)}, \qquad k_{2l}=2\pi l/L.
      \label{22} \end{equation}
Under periodic BCs, the potential can be expanded in a Fourier series
\begin{equation}
 U(|x_{1}-x_{2}|) = \frac{1}{L}
 \sum\limits_{j=0,\pm 1,\pm
2,\ldots}\nu(k_{j})e^{i k_{j}(x_{1}-x_{2})},
     \label{23} \end{equation}
 \begin{equation}
 \nu(k) = \int\limits_{-L}^{L} U_{1}(|x|)e^{-i kx}d x,
     \label{24} \end{equation}
where  $x=x_{1}-x_{2}$, and $k_{j}=2\pi j/L$. In this case, the
potential is
\begin{equation}
 U(|x_{1}-x_{2}|) = U_{1}(|x_{1}-x_{2}|)+U_{1}(L-|x_{1}-x_{2}|),
     \label{13d} \end{equation}
because one particle acts on the other particle from both sides.
Note that Eq. (\ref{24}) contains namely $U_{1}(x)$, rather than
$U(x)$. The formulae (\ref{23}) and (\ref{24}) follow from  Fourier
analysis if we consider that the argument(s) of the function
$U(|x_{1}-x_{2}|)$ is (are) [i]  $x_{1}-x_{2}$ or [ii] $x_{1}$ and
$x_{2}$ independently. In the thermodynamic limit, the addition
$U_{1}(L-|x_{1}-x_{2}|)$ into (\ref{13d}) is usually omitted. By
substituting (\ref{22}) and (\ref{23}) into  (\ref{20}) and
(\ref{21}), we can easily obtain the Bogoliubov dispersion law
(\ref{0}).

If we take boundaries into account and try to solve the problem in a
similar way,  we will fail.  This is due to the fact that, because
of the different expansion of the potential,  substituting the
harmonic (\ref{22}) into Eqs. (\ref{20}) and (\ref{21}) generates
many other harmonics. According to the rules of Fourier analysis,
for a system with zero BCs, the potential $U(|x_{1}-x_{2}|)$ can be
expanded into the following series:
\begin{eqnarray}
U(|x_{1}-x_{2}|) &=&  \sum\limits_{j=0,\pm 1,\pm
2,\ldots}\frac{\nu(k_{j})}{2L}e^{i k_{j}(x_{1}-x_{2})} \nonumber \\
&=&\frac{\nu(0)}{2L}+\sum\limits_{j=1,2,\ldots}\frac{\nu(k_{j})}{L}
\cos{[k_{j}(x_{1}-x_{2})]},
     \label{26} \end{eqnarray}
 \begin{equation}
 \nu(k) = \int\limits_{-L}^{L} U(|x|)e^{-i kx}d x,
     \label{27} \end{equation}
where $k_{j}=\pi j/L$. This series exactly reproduces the initial
function over the entire domain $x_{1},x_{2}\in\lbrack 0,L]$. The
expansions for periodic and zero BCs were analyzed in detail in
\cite{ryady}.

From equations (\ref{20}), (\ref{21}), and (\ref{26}), we now find
the dispersion law under  zero BCs
\begin{equation}
 \tilde{n}(0)=0, \quad \tilde{n}(L)=0.
\label{27bc} \end{equation}%
We represent  $\tilde{n}(x)$ and  $s(x)$ in the form of expansions
in the complete set of  sines:
\begin{equation}
\tilde{n}(x)=n_{0}\sum\limits_{l=1,2,\ldots}a_{l}\sin{(k_{l}x)},
      \label{28a} \end{equation}
\begin{equation}
 s(x)=\sum\limits_{l=1,2,\ldots}b_{l}\sin{(k_{l}x)},
      \label{28b} \end{equation}
where $k_{l}=\pi l/L$. Then the BCs (\ref{27bc}) are satisfied, and
the expansions (\ref{28a}) and (\ref{28b}) contain all possible
wavelengths ($\lambda=2L/l$) corresponding to the zero BCs.
Therefore, the expansions (\ref{28a}) and (\ref{28b}) ensure a
correct and complete description of the system.

Let us substitute these series into Eq. (\ref{20}). Because the
sines $\sin{(k_{l}x)}$ are independent basis functions, the
equations for the coefficients $a_{l}$, $b_{l}$ and frequency
$\omega$ can be obtained if the sum of the coefficients of each of
$\sin{(k_{l}x)}$ is set to zero. In this way, equation (\ref{20})
yields
\begin{equation}
b_{l}=-\frac{\omega m}{k_{l}^{2}}a_{l}, \qquad l=1,2,\ldots,\infty.
      \label{30} \end{equation}
Using  (\ref{26}), (\ref{28a}), and (\ref{28b}) and the expansions
\begin{equation}
\cos(k_{j}x)=\sum\limits_{p=1,2,3,\ldots}c_{j}^{p}\sin(k_{p}x),
\label{31a} \end{equation}
\begin{equation}
1=\sum\limits_{p=1,2,3,\ldots}c_{0}^{p}\sin(k_{p}x) \label{31b}
\end{equation}
with
\begin{equation} c_{j}^{p} = \left [
\begin{array}{ccc}
0  & \   \mbox{for even} \ p-j,   & \\
\frac{2}{\pi}\left (\frac{1}{p-j}+\frac{1}{p+j}\right )  & \mbox{for
odd} \ p-j, & \end{array} \right.   \label{31c}     \end{equation}%
we find after some algebra:
\begin{eqnarray}
&&\int\limits_{0}^{L} d
x^{\prime}\tilde{n}(x^{\prime})U(|x-x^{\prime}|)=
\frac{n_{0}\nu(0)}{2L}\sum\limits_{p=1,2,\ldots}\frac{a_{p}}{k_{p}}[1-(-1)^{p}]\sum\limits_{l=1,2,\ldots}c_{0}^{l}\sin{(k_{l}x)}
 \nonumber
\\ &&+\sum\limits_{l=1,2,\ldots}\sin{(k_{l}x)}\left
\{\frac{a_{l}n_{0}\nu(k_{l})}{2} \right.  \label{31} \\ &&+ \left.
\sum\limits_{j,p=1,2,\ldots}^{j\neq
p}\frac{c_{j}^{l}a_{p}n_{0}\nu(k_{j})}{2\pi} [1-(-1)^{p+j}]\left
(\frac{1}{p-j}+\frac{1}{p+j}\right )\right \}.
     \nonumber \end{eqnarray}
Subject to (\ref{28a}), (\ref{28b}), (\ref{30}), and (\ref{31}),
equation (\ref{21}) is reduced to:
\begin{eqnarray}
&&
\sum\limits_{l=1,2,\ldots}a_{l}\sin{(k_{l}x)}\frac{E^{2}_{M}(k_{l})-\hbar^{2}\omega^{2}}{\hbar^{2}k_{l}^{2}/m}
\nonumber
\\ && +\sum\limits_{l=1,2,\ldots}c_{0}^{l}\sin{(k_{l}x)}\sum\limits_{p=1,2,\ldots}\frac{n_{0}\nu(0)}{2\pi
p}a_{p}[1-(-1)^{p}] \label{32}
\\ && + \sum\limits_{l=1,2,\ldots}\sin{(k_{l}x)}\sum\limits_{j,p=1,2,\ldots}^{j\neq p}\frac{c_{j}^{l}n_{0}\nu(k_{j})}{2\pi}
a_{p}[1-(-1)^{p+j}]\left (\frac{1}{p-j}+\frac{1}{p+j}\right )=0,
     \nonumber \end{eqnarray}
 \begin{equation}
 E^{2}_{M}(k) = \left (\frac{\hbar^{2} k^2}{2m}\right )^{2} +
  n_{0}\nu(k)\frac{\hbar^2 k^2}{2m}.
      \label{Ef} \end{equation}
Because the functions $\sin{(k_{l}x)}$  are independent, from
(\ref{32}) we obtain the system of equations for the unknown
coefficients $a_{l}$ and  frequency $\omega$:
\begin{eqnarray}
&& (E^{2}_{M}(k_{l})-\hbar^{2}\omega^{2})a_{l}
+c_{0}^{l}\sum\limits_{p=1,2,\ldots}\frac{\epsilon_{l}^{2}(q_{0})}{\pi
p}[1-(-1)^{p}]a_{p}  \label{33}
\\ && + \sum\limits_{j,p=1,2,\ldots}^{j\neq p}\frac{c_{j}^{l}\epsilon_{l}^{2}(q_{j})}{\pi}
[1-(-1)^{p+j}]\left (\frac{1}{p-j}+\frac{1}{p+j}\right )a_{p}=0,
\quad l=1,2,\ldots, \infty,
     \nonumber \end{eqnarray}
where
\begin{equation}
\epsilon_{l}^{2}(q_{j})=n_{0}\nu(q_{j})\frac{\hbar^{2}k_{l}^{2}}{2m}.
      \label{35} \end{equation}
Because the factor $c_{0}^{l}$ is zero for all even $l$'s and
non-zero for all odd $l$'s, the system of equations (\ref{33})
splits into two independent systems of equations,  one for $a_{l}$'s
with even $l$'s and the other for $a_{l}$'s with odd $l$'s. These
systems of equations can be written as follows:
\begin{eqnarray}
&&a_{2l}(E^{2}_{M}(k_{2l})-\hbar^{2}\omega^{2})
+\frac{4}{\pi^{2}}\sum\limits_{j=0,\pm 1, \pm2,
\ldots}\frac{\epsilon_{2l}^{2}(k_{2j+1})}{2l-2j-1} \nonumber
\\ && \times\sum\limits_{p =1,2,\ldots}a_{2p}\left
[\frac{1}{2p-2j-1}+\frac{1}{2p+2j+1}\right ]=0,  \quad l=1,2,\ldots,
\infty,
      \label{40} \end{eqnarray}
\begin{eqnarray}
&&a_{2l-1}(E^{2}_{M}(k_{2l-1})-\hbar^{2}\omega^{2})
+\frac{4}{\pi^{2}}\sum\limits_{j=0,\pm 1, \pm2,
\ldots}\frac{\epsilon_{2l-1}^{2}(k_{2j})}{2l-2j-1}\nonumber
\\ && \times \sum\limits_{p
=1,2,\ldots}a_{2p-1}\left [\frac{1}{2p-1-2j}+\frac{1}{2p-1+2j}\right
]=0,   \quad l=1,2,\ldots, \infty.
      \label{40odd} \end{eqnarray}

We now take into account that, for any integer $l,$
\begin{equation}
\frac{4}{\pi^{2}}\sum\limits_{j=0, \pm 1, \pm 2,
\ldots}\frac{1}{(2l-2j-1)^{2}}=1.
      \label{41} \end{equation}
For a large $N$ the function $\epsilon_{2l}^{2}(q_{2j+1})$ varies
slightly as $j$ changes by one. The main contribution to  sum
(\ref{41}) is made by the terms with $j=l, l-1$ and the nearest
ones. Therefore, we have
\begin{equation}
\frac{4}{\pi^{2}}\sum\limits_{j=0, \pm 1, \pm 2,
\ldots}\frac{\epsilon_{2l}^{2}(k_{2j+1})}{(2l-2j-1)^{2}}=
\epsilon_{2l}^{2}(k_{2l})+\epsilon_{g}^{2}(2l),
      \label{42} \end{equation}
where $\epsilon_{g}^{2}(2l)\ll \epsilon_{2l}^{2}(k_{2l})$ should
hold. Eq. (\ref{42}) defines function $\epsilon_{g}^{2}(2l)$. Let us
separate the term with $p=l$ from the sum in (\ref{40}). Then,
subject to (\ref{42}), we obtain the following infinite system of
equations for the coefficients $a_{2l}$:
\begin{eqnarray}
&&a_{2l}(E^{2}_{B}(k_{2l})+\epsilon^{2}_{b}(2l)+\epsilon_{g}^{2}(2l)-\hbar^{2}\omega^{2})
+\frac{4}{\pi^{2}}\sum\limits_{j =0, \pm 1, \pm 2,
\ldots}\frac{\epsilon_{2l}^{2}(k_{2j+1})}{2l-2j-1} \nonumber
\\&&\times \sum\limits^{p\neq l}_{p =1, 2, \ldots}a_{2p}\left
[\frac{1}{2p-2j-1}+\frac{1}{2p+2j+1}\right ]=0,  \quad l=1,2,\ldots,
\infty,
      \label{46} \end{eqnarray}
with
\begin{equation}
\epsilon^{2}_{b}(2l)=\frac{4}{\pi^{2}}\sum\limits_{j =0, \pm 1, \pm
2, \ldots}\frac{\epsilon_{2l}^{2}(k_{2j+1})}{4l^{2}-(2j+1)^{2}}.
      \label{47} \end{equation}

Similarly, Eq. (\ref{40odd}) can be written as
\begin{eqnarray}
&&a_{2l-1}(E^{2}_{B}(k_{2l-1})+\epsilon^{2}_{b}(2l-1)+\epsilon_{g}^{2}(2l-1)-\hbar^{2}\omega^{2})
 \label{46odd}
\\&& +\frac{4}{\pi^{2}}\sum\limits_{j =0, \pm 1, \pm 2,
\ldots}\frac{\epsilon_{2l-1}^{2}(k_{2j})}{2l-2j-1}
\sum\limits^{p\neq l}_{p =1, 2, \ldots}a_{2p-1}\left
[\frac{1}{2p-1-2j}+\frac{1}{2p-1+2j}\right ]=0,
      \nonumber \end{eqnarray}
where $l=1,2,\ldots,\infty $ and
\begin{equation}
\epsilon^{2}_{b}(2l-1)=\frac{4}{\pi^{2}}\sum\limits_{j =0, \pm 1,
\pm 2, \ldots}\frac{\epsilon_{2l-1}^{2}(k_{2j})}{(2l-1)^{2}-4j^{2}},
      \label{47odd} \end{equation}
\begin{eqnarray}
\epsilon_{g}^{2}(2l-1)=\frac{4}{\pi^{2}}\sum\limits_{j=0, \pm 1, \pm
2, \ldots}\frac{\epsilon_{2l-1}^{2}(k_{2j})}{(2l-2j-1)^{2}}-
\epsilon_{2l-1}^{2}(k_{2l-1}).
      \label{42odd} \end{eqnarray}

It is convenient to introduce the coefficients $f_{l}$:
\begin{eqnarray}
f_{l}=\frac{\epsilon^{2}_{b}(l)+\epsilon_{g}^{2}(l)}{E^{2}_{B}(k_{l})},
\quad l=1,2,\ldots, \infty.
      \label{ff} \end{eqnarray}

Consider the system  of equations (\ref{46}) for $a_{2l}$'s. This is
an infinite system of linear homogeneous equations for the
coefficients $a_{2l}$. The system has a solution if its determinant
is equal to zero. This condition yields the equation for
$\omega^{2}$, which has an infinite number of solutions
$\omega^{2}_{2j}$ ($j=1, 2, \ldots, \infty$). The sum in Eq.
(\ref{46}) contains only  terms with alternating denominators. This
favors the smallness of the sum, because
\begin{equation}
\sum\limits_{j =0, \pm 1, \pm 2, \ldots}\frac{1}{(2l-2j-1)(\pm
2p-2j-1)}=0
      \label{48} \end{equation}
for integers $p$, $l$, and $\pm p \neq l$. These properties allow us
to construct a perturbation theory to solve equation (\ref{46}).

To obtain the solution  in the zero approximation, we set the sum in
(\ref{46}) to zero. Then the system  of equations (\ref{46}) is
reduced to the equations
\begin{equation}
a_{2l}(E^{2}_{B}(k_{2l})[1+f_{2l}]-\hbar^{2}\omega^{2})=0, \quad
l=1,2,\ldots, \infty
      \label{zero-1} \end{equation}
and gives the solutions $a_{2l_{0}}\neq 0$, $a_{2l\neq 2l_{0}}=0$
and
\begin{equation}
\hbar^{2}\omega^{2}(k_{2l_{0}})=E^{2}_{B}(k_{2l_{0}})[1+f_{2l_{0}}];
      \label{zero-2} \end{equation}
here $l_{0}, l=1, 2, \ldots, \infty$.   That is,  in the zero
approximation,   expansions (\ref{28a}) and (\ref{28b})  consist
only of a single $2l_{0}$-harmonic. In the higher approximations, we
will include other harmonics.

In the first approximation, we take into account the sum in
(\ref{46}) and consider that $a_{2l\neq 2l_{0}}$ are small but
non-zero for all $l, l_{0}$. Let $\omega(2l_{0})$ be the frequency,
which is a solution of the system  of equations (\ref{46}) and
corresponds to the quasimomentum $k_{2l_{0}}$: $\omega(2l_{0})\equiv
\omega(k_{2l_{0}})$. Then, in the set of coefficients $a_{2}, a_{4},
a_{6}, \ldots $, which are the solutions of Eq. (\ref{46}) for the
frequency $\omega(2l_{0})$, the value of the harmonic $a_{2l_{0}}$
must be much larger (in modulus) than the values of all the other
$a_{2l}$'s. Therefore, for the frequency $\omega(2l_{0})$ we
consider the harmonic $a_{2l_{0}}$ to be the principal and separate
it from the sum in Eq. (\ref{46}) by assuming that the contribution
of the other harmonics $a_{2p\neq 2l_{0}}$ to the sum is small. From
the system of equations (\ref{46}), for $l \neq l_{0}$ we find
\begin{equation}
a_{2l\neq
2l_{0}}(2l_{0})=a_{2l_{0}}(2l_{0})\frac{A_{2l}(2l_{0})\hbar^{2}\omega^{2}(2l_{0})}{E^{2}_{B}(k_{2l})[1+f_{2l}]
-\hbar^{2}\omega^{2}(2l_{0})},
      \label{49} \end{equation}
where $l=1,2,\ldots, \infty$ (except $l= l_{0}$), and the inequality
$|a_{2l\neq 2l_{0}}(2l_{0})|\ll |a_{2l_{0}}(2l_{0})|$ must hold
(otherwise the exact solution $\hbar^{2}\omega^{2}(k_{2l_{0}})$ can
be very different from the zero approximation (\ref{zero-2})). If
$|a_{2l\neq 2l_{0}}(2l_{0})|\ll |a_{2l_{0}}(2l_{0})|$, we can look
for the quantity $A_{2l}(2l_{0})$ using perturbation theory:
\begin{equation}
A_{2l}(2l_{0})=A_{2l}^{(0)}(2l_{0})+\delta A_{2l}(2l_{0}),
      \label{50} \end{equation}
\begin{eqnarray}
A_{2l}^{(0)}(2l_{0})&=&-\frac{4}{\pi^{2}}\sum\limits_{j =0, \pm 1,
\ldots, \pm \infty
}\frac{\epsilon_{2l}^{2}(k_{2j+1})}{\hbar^{2}\omega^{2}(2l_{0})}
\frac{1}{2l-2j-1}\nonumber \\&\times &\left
[\frac{1}{2l_{0}-2j-1}+\frac{1}{2l_{0}+2j+1}\right ],
      \label{51} \end{eqnarray}
\begin{eqnarray}
&&\delta A_{2l}(2l_{0}) = -\frac{4}{\pi^{2}}\sum\limits_{j =0, \pm
1, \ldots, \pm \infty}\frac{\epsilon_{2l}^{2}(k_{2j+1})}{2l-2j-1}
\nonumber \\&&\times  \sum\limits_{p=1,2,\ldots,\infty}^{p\neq
 l,  l_{0}}
\frac{A_{2p}(2l_{0})}{E^{2}_{B}(k_{2p})[1+f_{2p}]-\hbar^{2}\omega^{2}(2l_{0})}
\left [\frac{1}{2p-2j-1}+\frac{1}{2p+2j+1}\right ] \nonumber \\
&& \approx -\frac{4}{\pi^{2}}\sum\limits_{j =0, \pm 1, \ldots, \pm
\infty}\frac{\epsilon_{2l}^{2}(k_{2j+1})}{2l-2j-1} \label{52}
\\&&\times \sum\limits_{p=1,2,\ldots,\infty}^{p\neq
 l,  l_{0}}
\frac{A_{2p}^{(0)}(2l_{0})}{E^{2}_{B}(k_{2p})[1+f_{2p}]-\hbar^{2}\omega^{2}(2l_{0})}
\left [\frac{1}{2p-2j-1}+\frac{1}{2p+2j+1}\right ].
      \nonumber \end{eqnarray}
This method is true if for all $l_{0}$ and $l$ (except $l=l_{0}$),
\begin{equation}
|\delta A_{2l}(2l_{0})|\ll |A_{2l}^{(0)}(2l_{0})|.
      \label{53} \end{equation}

\begin{figure*}
\includegraphics[width=.6\textwidth]{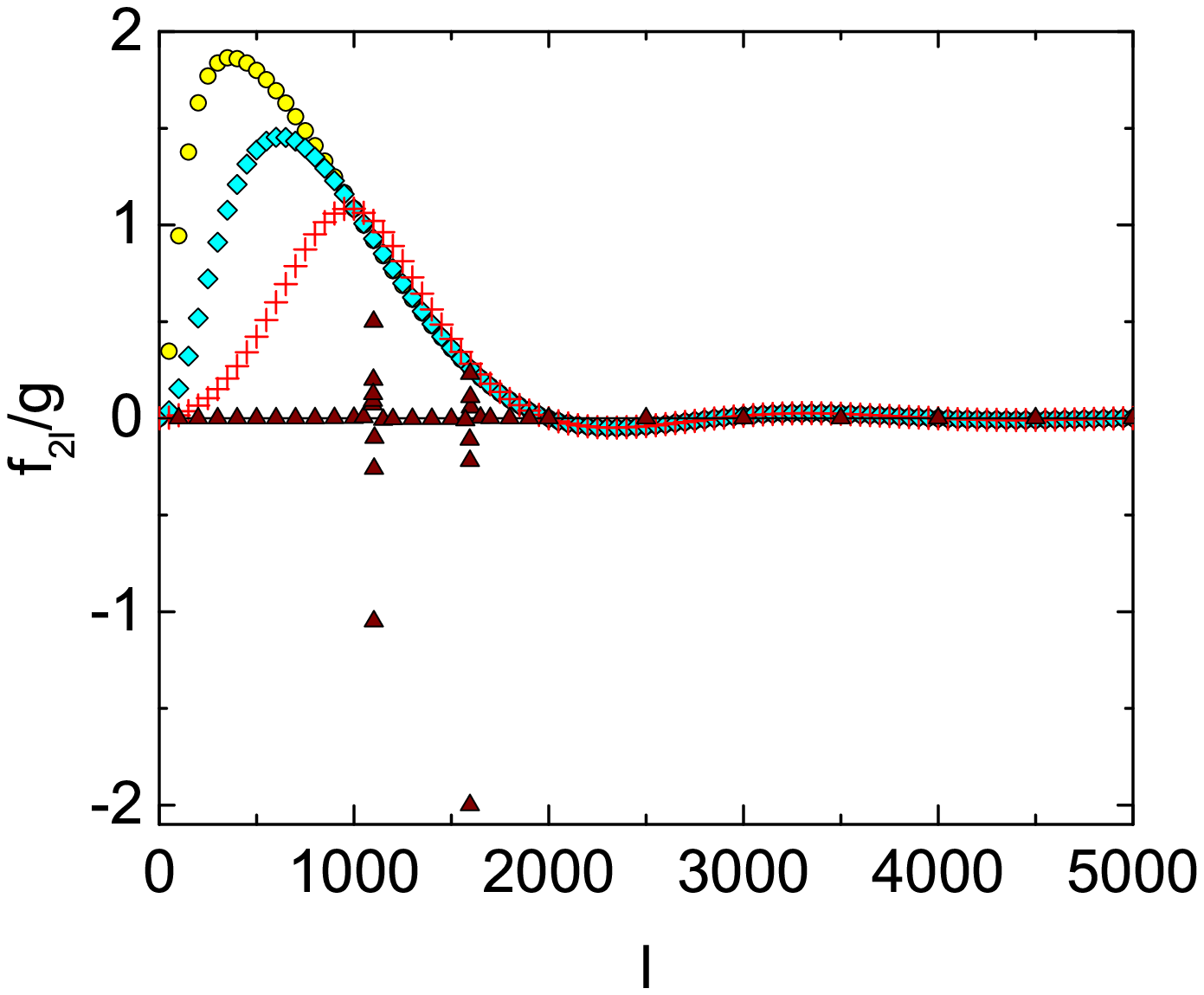}
\caption{[Color online]  Values of $f_{2l}/g$ for various $l$ for
the 1D system of $N=1000$ $^4$He atoms  with interatomic potential
(\ref{p1}) and $a= 2\,\mbox{\AA}$, $a/\bar{R}=0.5$.  The symbols
code the different values of $u_{0}=\frac{U_{0}}{1\mathrm{K}\cdot
k_{B}}$: $u_{0}=0.1$ (circles), $u_{0}=1$ (diamonds), $u_{0}=10$
(crosses), and $u_{0}=100$ (triangles). The scale factor $g$ is
different for different curves:
$g=0.1\frac{u_{0}}{N}\frac{a^{2}}{\bar{R}^{2}}$ for $u_{0}=0.1; 1;
10$ and $g=1$ for $u_{0}=100$. Note that we have changed the value
of $|f_{2l}/g|$ for $l=1596$ (the bottom triangle): the figure shows
$f_{2l}/g=-2$, but the real value is $f_{2l}/g=-6.35$. }
          \label{fig:1}                  \end{figure*}

Next, we consider Eq. (\ref{46})   with  $l= l_{0}$ and substitute
$a_{2l\neq 2l_{0}}$ (\ref{49}) into  (\ref{46}). In this case,
$a_{2l_{0}}$ is canceled in Eq. (\ref{46}), and we obtain the
following equation for the frequency:
\begin{equation}
\hbar^{2}\omega^{2}(2l_{0})=E^{2}_{B}(k_{2l_{0}})[1+f_{2l_{0}}]
-\hbar^{2}\omega^{2}(2l_{0})\delta A_{2l_{0}}(2l_{0}).
      \label{54} \end{equation}
Although equation (\ref{52}) was obtained for $\delta
A_{2l}(2l_{0})$ when $l \neq l_0$, we may formally set $l=l_0$ in
(\ref{52}). Such a $\delta A_{2l}(2l_{0})$ is denoted in (\ref{54})
as $\delta A_{2l_{0}}(2l_{0})$. Let us also denote
\begin{equation}
-\hbar^{2}\omega^{2}(2l_{0})\delta A_{2l_{0}}(2l_{0}) =
q_{2l_{0}}E^{2}_{B}(k_{2l_{0}}),
      \label{55-3} \end{equation}
then (\ref{54}) takes the form
\begin{equation}
\hbar^{2}\omega^{2}(2l_{0})=E^{2}_{B}(k_{2l_{0}})[1+f_{2l_{0}}+q_{2l_{0}}].
      \label{54-2} \end{equation}
From Eqs. (\ref{55-3}) and  (\ref{54-2})  we get the equation
\begin{equation}
q_{2l_{0}}= -\delta A_{2l_{0}}(2l_{0})[1+f_{2l_{0}}+q_{2l_{0}}]
      \label{59-2} \end{equation}
for the new parameter $q_{2l_{0}}$. In this case, $\delta
A_{2l_{0}}(2l_{0})$ is given by (\ref{52}) and depends on
$q_{2l_{0}}$.

\begin{figure*}
\includegraphics[width=.6\textwidth]{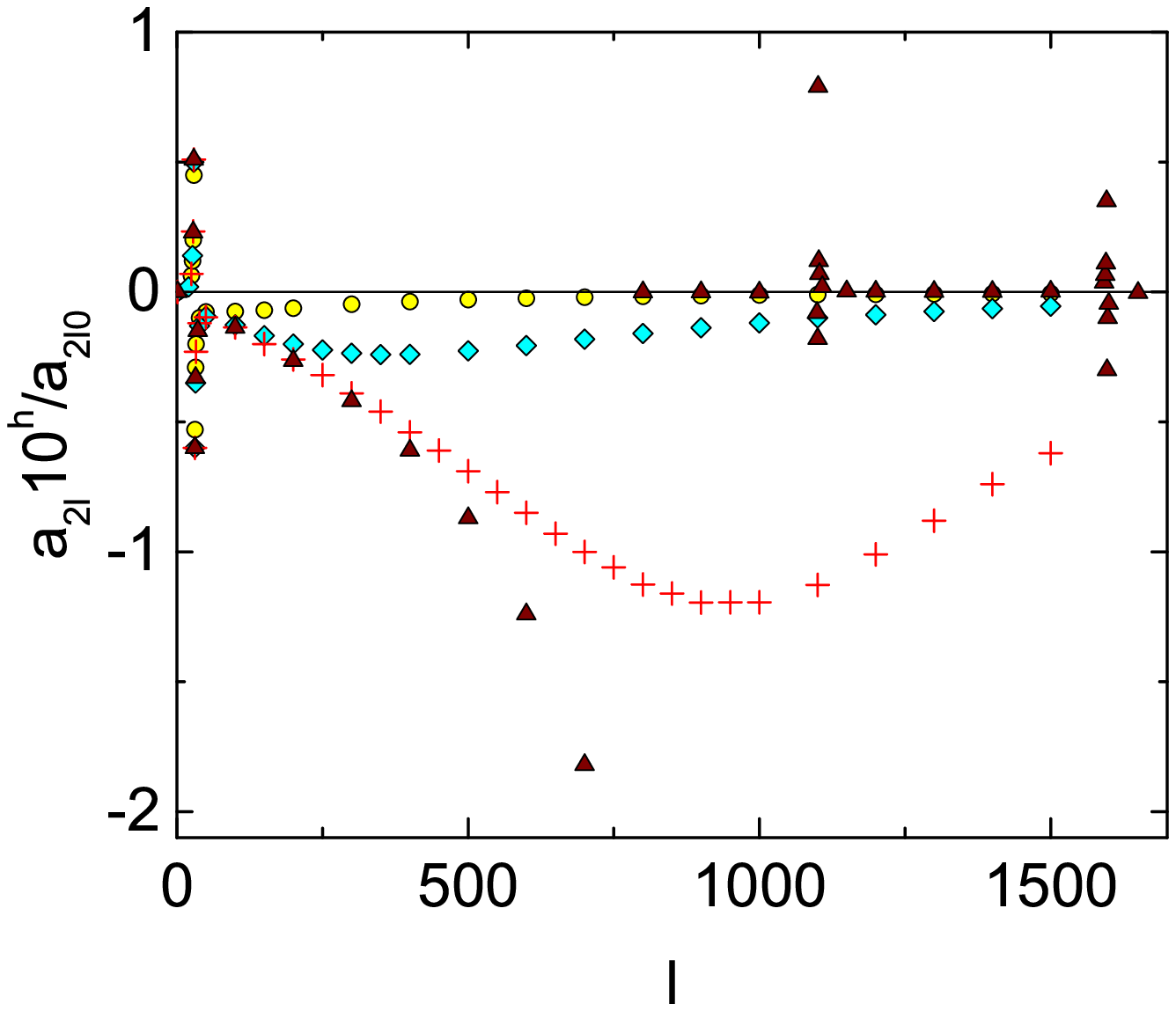}
\caption{[Color online]  The set of quantities
$a_{2l}(2l_{0})/a_{2l_{0}}(2l_{0})$ for wave packet $\tilde{n}(x)$
(Eqs. (\ref{28a}), (\ref{28b})) for the $30$th frequency
($l_{0}=30$). The $a_{2l}$ are obtained by the formula (\ref{49}),
where $A_{2l}(2l_{0})=A_{2l}^{(0)}(2l_{0})$ and
$A_{2l}^{(0)}(2l_{0})$ is given by (\ref{51}).    Shown are the
curves for $u_{0}=0.1$ (circles), $1$ (diamonds), $10$ (crosses),
and $100$ (triangles). We consider $N=1000$ $^4$He atoms with the
interatomic potential (\ref{p1}) for $a= 2\,\mbox{\AA}$,
$a/\bar{R}=0.5$ and $q_{2l_{0}}=0$. The values of $a_{2l}(2l_{0})$
are multiplied by a factor $10^h$, which is equal to $10^5$ for
$u_{0}=0.1; 1; 10$. For $u_{0}=100$ we assume $10^{h}=10^{5}$ when
$l\leq 750$ and $10^{h}=10$ when $l>750$. The discontinuity in the
curve $\triangle\triangle\triangle$ at $l= 750$ is fictitious and is
only caused by the change in the coefficient $10^{h}$. }
          \label{fig:2}                  \end{figure*}

It is difficult to analyze the obtained equations analytically, but
the numerical analysis is rather simple. We used the
``semi-transparent sphere'' potential
\begin{equation}
 U(|x_{1}-x_{2}|) =
\left [ \begin{array}{ccc}
    U_{0}>0,  & \    |x_{1}-x_{2}|\leq a  & \\
        0,  & \ a <|x_{1}-x_{2}|\leq L &
\label{p1} \end{array} \right. \end{equation}%
for the domain $x_{1}, x_{2}\in [0,L]$ and solved the system of
equations (\ref{46}) for $l=1,2,\ldots,l^{max}$ ($l^{max}\simeq 10 N
\bar{R}/a$) and several values of $a/\bar{R}$ and
$u_{0}=\frac{U_{0}}{1\mathrm{K}\cdot k_{B}}$ using perturbation
theory (\ref{49})--(\ref{59-2}). The summations in (\ref{42}),
(\ref{47}), (\ref{51}), and (\ref{52}) were performed over
$p=1,2,\ldots,l^{max}$ and $j=0,\pm 1,\ldots,\pm l^{max}$;
increasing $l^{max}$ by a factor of $10$ changed the results
negligibly. For $^4$He atoms ($a \approx 2\,\mbox{\AA}$, $m\approx
6.65 \cdot 10^{-24}$g) the numerical analysis shows that the numbers
$f_{2l}$ (\ref{ff}) are very small ($|f_{2l}|\lsim 1/N$, see Fig. 1)
when $u_{0}a/\bar{R}\lsim 10$. In this case, the largest among
$|f_{2l}|$'s is equal to $|f_{2l}|\simeq
\frac{0.1u_{0}}{N}\frac{a^p}{\bar{R}^p}$ with $p\approx 1$--$2$. The
inequalities $|a_{2l\neq 2l_{0}}(2l_{0})|\ll |a_{2l_{0}}(2l_{0})|$
and (\ref{53}) are satisfied provided that $u_{0}a/\bar{R}\lsim 1$.
The values of $a_{2l}(2l_{0})/a_{2l_{0}}(2l_{0})$ and $\delta
A_{2l}(2l_{0})/A^{(0)}_{2l}(2l_{0})$ for the different sets of
parameters are shown in Figs. 2 and 3. When $u_{0}a/\bar{R}\gg 1$,
several values $a_{2l}(2l_{0})/a_{2l_{0}}(2l_{0})$ and all $\delta
A_{2l}(2l_{0})/A^{(0)}_{2l}(2l_{0})$ are of the order of unity (in
modulus); therefore, our perturbation theory does not work. We
verified this for $N=100, 1000, 10000$ and many values of $l_{0}$.

In the numerical analysis, we used $\hbar^{2}\omega^{2}$
(\ref{54-2}) with $q$ as a free parameter. We varied $q$ from -100
to 100 and compared it with the theoretical $q$ which is given by
the left-hand side of (\ref{59-2}). The solution for $q$ is that $q$
for which the theoretical $q$ is equal to the free $q$.  We obtained
$|q|\lsim 0.01/N$ for $u_{0}a/\bar{R}\lsim 1$, $N= 100, 1000$  and
$l_{0}=1,2,\ldots,l^{max}$ with $l^{max}\simeq 10 N \bar{R}/a$. Note
that (\ref{59-2}) is an algebraic equation of infinite degree with
respect to $q_{2l_{0}}$. Therefore, there must be an infinite number
of roots $q_{2l_{0}}$. All roots, except $q_{2l_{0}} =0$ are
probably complex.

\begin{figure*}
\includegraphics[width=.6\textwidth]{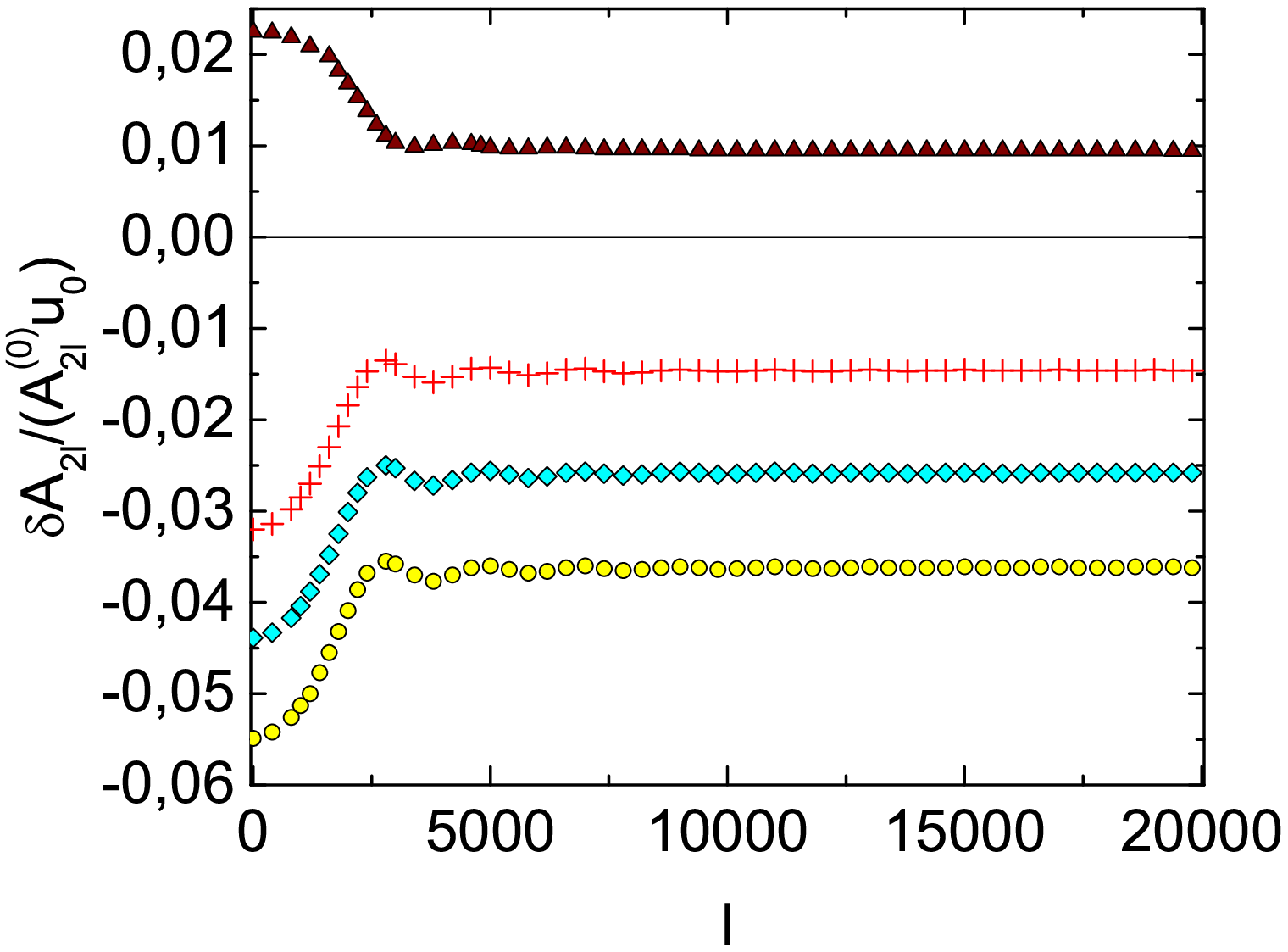}
\caption{[Color online]  Values of $\delta
A_{2l}(2l_{0})/[A^{(0)}_{2l}(2l_{0})u_{0}]$ for various $l$'s for
the system of $N=1000$ $^4$He atoms  with interatomic potential
(\ref{p1}); $l_{0}=30$, $q_{2l_{0}}=0$, $a= 2\,\mbox{\AA}$, and
$a/\bar{R}=0.5$. Shown are the curves for $u_{0}=0.1$ (circles), $1$
(diamonds), $10$ (crosses), and $100$ (triangles).  The values of
$A^{(0)}_{2l}(2l_{0})$ and $\delta A_{2l}(2l_{0})$ are obtained by
the formulae (\ref{51}), (\ref{52}). }
          \label{fig:3}                  \end{figure*}

Thus, the numerical analysis shows that $|f_{2l}|\lsim 1/N$ and
$|q(2l_{0})|\lsim 0.01/N$ for all $l$ and $l_{0}$ provided that
$u_{0}a/\bar{R}\lsim 1$. In this case, Eq. (\ref{54-2}) takes the
form
\begin{equation}
\hbar^{2}\omega^{2}(k)= E^{2}_{B}(k),
      \label{60} \end{equation}
which is equivalent to the famous Bogoliubov formula (\ref{0}).

An analysis of the system of equations (\ref{46odd}) for odd
harmonics also leads to the Bogoliubov dispersion law. In this case,
the formulae and figures are similar; therefore, we skip them.

As mentioned in the Introduction, one of the difficulties in the
case of zero BCs is determining the quasimomentum of the
quasiparticle. The above analysis allows one to find quasimomentum.
Because $|a_{2l}(2l_{0})/a_{2l_{0}}(2l_{0})|\ll 1$ for all $l,
l_{0}=1,2,\ldots,l^{max}$ (for $l\neq l_{0}$),  harmonic
$a_{2l_{0}}$ strongly dominates  the wave packet $\tilde{n}(x)$
(\ref{28a}). Therefore,  $k_{2l_{0}}=\pi 2l_{0}/L$ is the
quasimomentum of the quasiparticle. Similarly, one can obtain that
for an odd harmonic, the quasimomentum is $k_{2l_{0}-1}=\pi
(2l_{0}-1)/L$. Thus, we have a general formula for the quasiparticle
quasimomentum,  $k_{l_{0}}=\pi l_{0}/L$ ($l_{0}=1,2,\ldots,\infty$),
which agrees with the results of papers
\cite{mtsp2019,mtmethodbog,cazalilla2004,cazalilla2002}.

\section{Discussion}
 The question arises: Have we found all solutions for
the frequencies? The answer is yes. Indeed,  system (\ref{46}) can
be solved for a finite number of $l$: $l=1, 2, \ldots, l^{max}$. By
equating the corresponding determinant to zero, we obtain an
algebraic equation of degree $l$ for $\omega^{2}$. Hence, the
$l^{max}$ solutions must exist. But we found exactly $l^{max}$
values of $\omega^{2}(2l_{0})$, which correspond to $l_{0}=1, 2,
\ldots, l^{max}$ and quasimomenta $k_{2}, k_{4},\ldots,
k_{2l^{max}}$. So there are no other solutions. Similarly for system
(\ref{46odd}). We can look at this from the other side. Condition
(\ref{zero-2}) virtually numbers the frequencies $\omega^{2}$;
therefore, the equality
$E^{2}_{B}(k_{2p})[1+f_{2p}]=\hbar^{2}\omega^{2}(2l_{0})$ is only
possible if $p= l_{0}$. This prohibits zero denominators
$E^{2}_{B}(k_{2p})[1+f_{2p}]-\hbar^{2}\omega^{2}(2l_{0})$ in
(\ref{52}) and hence large values of $|\delta A_{2l_{0}}(2l_{0})|$
in (\ref{59-2}) at $u_{0}a/\bar{R}\lsim 1$. Therefore, the singular
denominators in (\ref{52}) do not lead to additional  solutions for
$q_{2l_{0}}$'s. Similarly for  $q_{2l_{0}+1}$'s.

Note that the exact solution for the ground-state energy $E_{0}$ of
a 1D system of point bosons
($U(|x_{j}-x_{l}|)=2c\delta(x_{j}-x_{l})$) is close to Bogoliubov's
$E_{0}$ only for $N\gsim 1000$ \cite{gp1,mt2015}. Consequently,
since Gross' approximation $\hat{\Psi}(\textbf{r},t)  =
\Psi(\textbf{r},t)$ \cite{gross1957} is in fact the zero
approximation of Bogoliubov's approach \cite{bog1947},  Gross'
equation (\ref{1}) can be applied to systems with $N\gsim 1000$. The
criterion for the applicability of Bogoliubov's model in the 1D case
for zero BCs and the point potential
$U(|x_{j}-x_{l}|)=2c\delta(x_{j}-x_{l})$ is
$\frac{\sqrt{\gamma}}{2\pi}\ln{\frac{N\sqrt{\gamma}}{\pi}}\ll 1$ (at
$T=0$) \cite{mtmethodbog} where $\gamma=\frac{2mc}{\hbar^{2}n}$.
This criterion can be written as:
\begin{equation}
\gamma \ll b=\left
(\frac{2\pi}{\ln{\frac{N\sqrt{\gamma}}{\pi}}}\right )^{2}.
     \label{cond1} \end{equation}
For $^4$He atoms we have $a \approx 2\,\mbox{\AA}$ and $\hbar
^{2}/(2ma^{2})\approx 1.5k_{B}\,\mathrm{K}$, so the condition
(\ref{cond1}) is equivalent to $u_{0}\ll 1.5ba/\bar{R}$, since
$n=1/\bar{R}$ and $c=\nu(0)/2=U_{0}a$ for the potential (\ref{p1}).
For the system to be uniform far from the walls, the inequality
$N\sqrt{\gamma}\gg 1$ should be satisfied \cite{mtmethodbog,gp1}.
Therefore, we have $b\sim 1$ for a not too large $N$, and
$b\rightarrow 0$ for $N\rightarrow \infty$. The usability condition
$u_{0}a/\bar{R}\lsim 1$ of the perturbation theory constructed above
is equivalent to the condition $u_{0}\lsim \bar{R}/a$. For
real-world systems, $\bar{R}/a\geq 1$. According to these estimates,
the condition $u_{0}\ll 1.5ba/\bar{R}$ required for Bogoliubov's
approach to work, sets a narrower range of variability of $u_{0}$ as
compared to the range $u_{0}\lsim \bar{R}/a$, in which our
perturbation theory holds. Thus, our method works over the entire
range of parameters for which Bogoliubov's approach holds.

On the other hand, a direct comparison of solutions of the
Gross-Pitaevskii equation (\ref{4}) with exact Bethe-ansatz
solutions shows that Gross' equation describes well the Bose system
with the interatomic potential
$U(|x_{j}-x_{l}|)=2c\delta(x_{j}-x_{l})$ and $\gamma \lsim 0.1, N
\gg 1$ \cite{gp1}. The condition $\gamma \lsim 0.1$ can be written
for the potential (\ref{p1}) as $2m u_{0}\mathrm{K}\cdot k_{B}a
\lsim 0.1\hbar^{2}n$,  inasmuch as $\gamma=\frac{2mc}{\hbar^{2}n}$
and $c=\nu(0)/2=U_{0}a=u_{0}\mathrm{K}\cdot k_{B}a$. For $^4$He
atoms (see above) this condition gives $ u_{0}a/\bar{R} \lsim
0.15(a/\bar{R})^{2}$.  Since for real systems $a/\bar{R}<1$, the
condition $ u_{0}a/\bar{R} \lsim 0.15(a/\bar{R})^{2}$ sets a much
stronger restriction than the applicability condition
$u_{0}a/\bar{R}\lsim 1$ of the perturbation theory. In other words,
our perturbation theory works over the entire parameter range in
which Gross' equation holds.

Interestingly, Bogoliubov's \textit{solutions} \cite{bog1947,bz1955}
for $E_{0}$ and $E(k)$ are valid for a much wider range of
parameters than the scope (\ref{cond1}) of Bogoliubov's
\textit{method}: for a 1D system of point bosons, the exact
solutions for the ground-state energy $E_{0}$
\cite{gp1,mt2015,ll1963} and the dispersion law $E(k)$
\cite{lieb1963,mtsp2019,gp2} are close to Bogoliubov's solutions at
$\gamma \lsim 1, N \gg 1$ (even for $N\rightarrow \infty$, which is
inconsistent with condition (\ref{cond1})). The cause of this is
still unclear.

The form of the asymptotic law $E(k\rightarrow 0)$ for the Bose gas
is also of interest. In works
\cite{bog1947,bz1955,yuv2,kulish1976,pash2010} the asymptotic law
$E(k\rightarrow 0)=|c_{1}|k+|c_{3}|k^{3}+\ldots$ was  found. The
formula (\ref{60}) agrees with this law. In paper \cite{popov1972b}
the relation $E(k)=|c_{1}|k$ was obtained. However, Eqs. (3.3) and
(3.6) of this paper clearly lead to the formula $E(k\rightarrow
0)=|c_{1}|k+|c_{3}|k^{3}+\ldots$. Another result, $E(k\rightarrow
0)=|c_{1}|k+|c_{2}|k^{2}+\ldots$, was obtained in
\cite{ristivojevic2014,pustilnik2014}. We do not comment on these
results because we did not investigate this issue. Note only that in
papers \cite{lieb1963,ristivojevic2014,pustilnik2014}  a point-like
potential was considered,  and the solutions $E(k)$ were
additionally obtained for the so-called hole-like excitations, which
were not found in the models
\cite{bog1947,bz1955,yuv2,pash2010,popov1972b}. However, this does
not mean that the models
\cite{bog1947,bz1955,yuv2,pash2010,popov1972b} and our approach  do
not catch the hole-like excitations. According to the analysis in
\cite{gp2,holes2020}, when the coupling is weak, the hole-like
excitation is a collection of identical interacting phonons. In the
present case, because the phonons are Bogoliubov's quasiparticles,
only Bogoliubov's quasiparticles are elementary excitations.

     \section{Conclusion}
\label{sec:4} We have found the dispersion law of a one-dimensional
weakly interacting zero-temperature Bose gas under zero boundary
conditions by solving Gross' equation using  special perturbation
theory. Such a method is mainly analytical (numerical analysis is
only used to prove the smallness of certain quantities and to
determine the range of applicability of the method). Note that all
frequencies $\omega$ can be found numerically from the systems of
equations (\ref{46}) and (\ref{46odd}) by setting the determinants
of the two corresponding matrices to zero. This approach is mostly
numerical and is beyond the scope of this study.

Our analysis shows that the dispersion law of a Bose gas with zero
BCs coincides with the dispersion law of the same periodic system. A
similar result was previously obtained by other methods
\cite{mtsp2019,mtmethodbog,cazalilla2004,cazalilla2002}. According
to results in \cite{ceperley2001}, the ground-state energy of a
Fermi system is the same for periodic and twisted BCs.

\section*{Acknowledgements}
The author is grateful to Yu.~Shtanov for discussions.  This
research was supported by the National Academy of Sciences of
Ukraine (Project No.~0123U102283) and the Simons Foundation.



       \end{document}